\documentclass{aa} % for a paper on 1 column  
\usepackage{graphicx}
\usepackage{txfonts}
\usepackage{lscape} % for 'landscape' environment

\begin{document}

   \title{Rapid evolution of [WC] stars in the Magellanic Clouds\thanks{Reduced spectra are only available in electronic form at the CDS via anonymous ftp to cdsarc.u-strasbg.fr (130.79.128.5) or via http://cdsweb.u-strasbg.fr/cgi-bin/qcat?J/A+A/}}

   \author{Marcin Hajduk
          \inst{1}
          }

   \institute{Space Radio-Diagnostics Research Centre, University of Warmia and Mazury, ul.Oczapowskiego 2, 10-719 Olsztyn, Poland\\
              \email{marcin.hajduk@uwm.edu.pl}
             }

   \date{Received September 15, 1996; accepted March 16, 1997}

% \abstract{}{}{}{}{} 
% 5 {} token are mandatory
 
  \abstract{We obtained new spectra of fourteen Magellanic Cloud planetary nebulae with the South African Large Telescope to determine heating rates of their central stars and to verify evolutionary models of post-asymptotic giant branch stars. We compared new spectra with observations made in previous years. Five planetary nebulae showed an increase in excitation over time. Four of their central stars exhibit [WC] features in their spectra, including three new detections. This raises the total number of [WC] central stars of PNe in the Magellanic Clouds to ten. We compared determined heating rates of the four [WC] central stars with the He-burning post-asymptotic giant branch evolutionary tracks and the remaining star with the H-burning tracks. Determined heating rates are consistent with the evolutionary models for both H and He-burning post-asymptotic giant branch stars. The central stars of the PNe that show the fastest increase of excitation are also the most luminous in the sample. This indicates that [WC] central stars in the Magellanic Clouds evolve faster than H-burning central stars, and they originate from more massive progenitors.}


   \maketitle
%
%-------------------------------------------------------------------

\section{Introduction}

Stars with initial masses in the range of $1-8\,\mathrm{M}_{\odot}$ lose most of their envelope on the asymptotic giant branch (AGB). During an early AGB phase, helium burning proceeds in a thin shell on the top of the carbon-oxygen core. When the helium shell reaches the hydrogen shell, the nuclear burning becomes unstable. A series of helium shell flashes punctuate quiescent hydrogen burning. 

The large energy generation due to a helium flash induces a convective instability between the shells \citep{1983ARA&A..21..271I,2005ARA&A..43..435H}. When the H shell becomes active again, part of the intershell material is dredged up to the surface. 

The envelope of the star is rapidly expelled due to the intensive mass loss experienced during the AGB phase. This reduces the number of helium shell flashes. The mass ejected by AGB stars significantly contributes to the chemical evolution of galaxies and stellar systems \citep{2010MNRAS.403.1413K}.

After a star ejects most of its envelope, it terminates the AGB phase and its evolution accelerates. Heating rate increases from $\dot{T}_* {\lesssim} 0.1 \, \mathrm{kK \, yr^{-1}}$ to $1 \, \mathrm{K \, yr^{-1}} {\lesssim} \dot{T}_* {\lesssim} 10 \, \mathrm{kK \, yr^{-1}}$ \citep{2019IAUS..343...36M}. The evolutionary timescale of post-AGB stars is a very strong function of their envelope mass and the nuclear burning rate. Post-AGB stars occupy a relatively narrow range in mass between $0.5$ and $0.9\,\mathrm{M}_{\odot}$, but evolutionary timescales differ by a few orders of magnitude within this range.

Post-AGB stars cross the H-R diagram from low to high temperatures almost at constant luminosity, as long as the shell sources remain active \citep{1970AcA....20...47P}. Once the star is hot enough, the ionization front expands throughout the shell ejected during the AGB phase. The shell forms a planetary nebula (PN). The excitation of the PN increases with the increasing temperature of the central star.

A central star of a PN may experience a final thermal pulse after it leaves AGB \citep{1979A&A....79..108S}. A final thermal pulse can punctuate evolution of a post-AGB star immediately after it leaves the AGB (called AFTP), during its horizontal evolution, when H-burning shell is active (late thermal pulse - LTP), or during cooling (very late thermal pulse - VLTP) \citep{2001Ap&SS.275...41K}. 

In the result of the VLTP, the pulse driven convecting zone eventually reaches out into the H-rich envelope \citep{2005ARA&A..43..435H}. The LTP and AFTP lead to the diluted shell and significantly reduced H abundance. An AFTP can lead to both a considerable enrichment with carbon and oxygen and to the dilution of hydrogen. This depends on the envelope mass. The smaller envelope mass, the less hydrogen is left in the envelope \citep{2001Ap&SS.275....1B}.

H-deficient central stars constitute at least 30\% of all the central stars in the Galaxy \citep{2011A&A...526A...6W}. More than half of them are [WC] stars, which show emission line spectra typical for WR stars of carbon sequence \citep{2003A&A...403..659A}. Hydrogen-deficient central stars are believed to expose their intershell region in their atmospheres due to a late helium-shell flash \citep{2006PASP..118..183W}.

The evolution of post-AGB stars remains one of the most difficult topics in stellar evolution \citep{2019IAUS..343...36M}. The models must rely on assumptions and approximations such as mixing length theory or stellar wind theory. Observational verification is necessary for the validation of the evolutionary models.

\begin{table*}[h!]
\begin{center}
 \caption[]{Log of the PNe observations.\label{tab:salt}}
\begin{tabular}{lcccc}
 \hline \hline
  Name &
  Exp. [s] &
  Date &
  Airmass &
  Other obs.\\
 \hline
Jacoby LMC 17&  2703    &       2013 Nov 08     & 1.33  & [1],[2]  \\
MGPN LMC 35     &       2845    &       2013 Nov 09     & 1.21  & [2],[3]  \\
MGPN LMC 39     &       2103    &       2013 Dec 22     & 1.26  & [2],[3]  \\
SMP LMC 26      &       2703    &       2013 Dec 21     & 1.29  & [4]  \\
SMP LMC 31      &       2785    &       2013 Nov 09     & 1.30  & [2],[4],[5]  \\
SMP LMC 55      &       2703    &       2013 Dec 22     & 1.29  & [2],[4],[5]  \\
SMP LMC 64      &       2103    &       2013 Nov 07     & 1.25  & [2],[5],[6]  \\
SMP LMC 67      &       2733    &       2013 Dec 23     & 1.24  & [2],[5],[6],[7]  \\
SMP LMC 104A &  2703    &       2013 Nov 10     & 1.33  & [7]  \\
MGPN SMC 8      &       2703    &       2013 Nov 08     & 1.29  & [3],[4],[8],[9]  \\
SMP SMC 1       &       2103    &       2013 Nov 08     & 1.33  & [1],[5],[9]  \\
SMP SMC 12      &       2103    &       2013 Nov 07     & 1.36  & [7],[9]  \\
SMP SMC 16      &       2103    &       2013 Nov 09     & 1.35  & [1],[3],[5],[6]  \\
SMP SMC 28      &       2703    &       2013 Nov 10     & 1.35  & [4],[6],[7]  \\
\hline
\end{tabular}
\end{center}
\tablebib{[1] \citet{1989ApJ...339..844B},
[2] \citet{2006MNRAS.373..521R}
[3] \citet{1992ApJS...83...87V},
[4] \citet{1988MNRAS.234..583M},
[5] \citet{1991ApJS...75..407M},
[6] \citet{2006ApJS..167..201S},
[7] \citet{2006AA...456..451L},
[8] \citet{2010ApJ...717..562S},
[9] \citet{2003ApJ...596..997S}.
}
\end{table*}

Stellar luminosities and temperatures have been traditionally used to study the evolution of post-AGB stars. However, post-AGB stars occupy a narrow range in luminosities in the H-R diagram. Their distances are little known and the stars emit most of their energy in the ultraviolet range. Observational errors propagate large uncertainties in the determination of the luminosities.

An alternative set of parameters useful for studying the evolution of post-AGB stars is the stellar temperature and heating rate. The heating rate is a sensitive measure of the stellar mass. The pace of the temperature evolution is determined by the ratio of the stellar luminosity and remaining envelope mass. 

\citet{2014A&A...566A..48G} derived heating rates from stellar temperatures and nebular ages for a sample of Galactic Bulge PNe. They needed to accelerate the evolutionary models by a factor of three to fit the white dwarf mass distribution and asteroseismological masses of central stars of PNe.

New models by \citet{2016A&A...588A..25M} confirmed the results obtained by \citet{2014A&A...566A..48G}. He obtained three to ten shorter timescales of post-AGB evolution than \citet{1995A&A...299..755B} and \citet{1994ApJS...92..125V}. His models were also brighter by ${\sim}{0.1-0.3}$ dex for the same mass. He took advantage of an updated treatment of the constitutive microphysics and included an updated description of the mixing processes and winds. 

An alternative method to derive heating rate is to measure the temperature evolution of the central star. This could be derived indirectly by comparison of the nebular line fluxes in different epochs. The change of the [O\,{\sc iii}] 5007\,\AA-to-H\,$\beta$ flux ratio in time is sensitive to the temperature evolution of cool central stars \citep{2015A&A...573A..65H}.

In this paper, we derived heating rates for a sample of central stars in the Magellanic Clouds and compared them with the existing evolutionary models. An advantage of Magellanic Cloud objects over Galactic PNe is the precisely determined distance. Moreover, Magellanic Cloud PNe are often compact and spatially unresolved by ground based telescopes. In such cases, the ion stratification in the nebula would not affect the measured flux ratios.

%--------------------------------------------------------------------

\section{Observations}

\begin{table}
 \caption[]{Log of the standard star observations.\label{tab:std}}
\begin{center}
\begin{tabular}{lccc}
 \hline \hline
  Name &
  Exp. [s] &
  Date &
  Air mass \\
 \hline
HILT\,600       &       180     &       2013 Nov 07     & 1.23  \\
HILT\,600   &   180     &       2013 Nov 08     & 1.23  \\
HILT\,600       &       180     &       2013 Nov 09     & 1.22  \\
EG\,21          &       60      &       2013 Dec 22     & 1.24  \\
\hline
\end{tabular}
\end{center}
\end{table}

We observed the fourteen PNe with the South African Large Telescope (SALT) equipped with the Robert Stobie Spectrograph \citep{2003SPIE.4841.1634K,2003SPIE.4841.1463B}. The SALT telescope has a fixed azimuthal angle. The Magellanic Clouds were accessible with the SALT near the meridian (hour angle from $-2$ to 2), so that the observing track length was sufficiently long. Under these conditions, atmospheric refraction should not play a significant role.

Three chips recorded each spectrum in different wavelength ranges: 4345-5347\,\AA, 5403-6393\,\AA, and 6448-7397\,\AA. We used longslit with the projected width of two arcsec. The resulting resolution was about 800 in the center of the spectral range. The slit was oriented north-south. The range of air masses for the targets and standards was $1.21-1.36$ (Tables \ref{tab:salt} and \ref{tab:std}). The log of observations is given in Table \ref{tab:salt}.

The spectra were extracted and wavelength calibrated. We calibrated the flux scale using a standard observed during the same night. On three nights: November 10, December 21, and December 22, flux standards were not observed. For these three nights we used the sensitivity curve derived for the closest night.

We measured the emission line fluxes with the Gaussian fit. We computed the errors of individual lines according to the formulae provided by \citet{1992PASP..104.1104L}. We added the derived error (root mean-square) of the sensitivity curve. We computed an error using at least two closest sensitivity curves if the standard was not observed during the given night. The fluxes were dereddened using the extinction law given by \citet{1999PASP..111...63F} and the total to selective extinction ratio $R_V=3.1$ (Table \ref{tab:fluxes}). We derived the extinction from the observed $\mathrm{H}\alpha$-to-$\mathrm{H}\beta$ line flux ratio.

\section{New [WC] central stars}

   \begin{figure*}[h!]
   \centering
   \includegraphics[width=0.9\hsize]{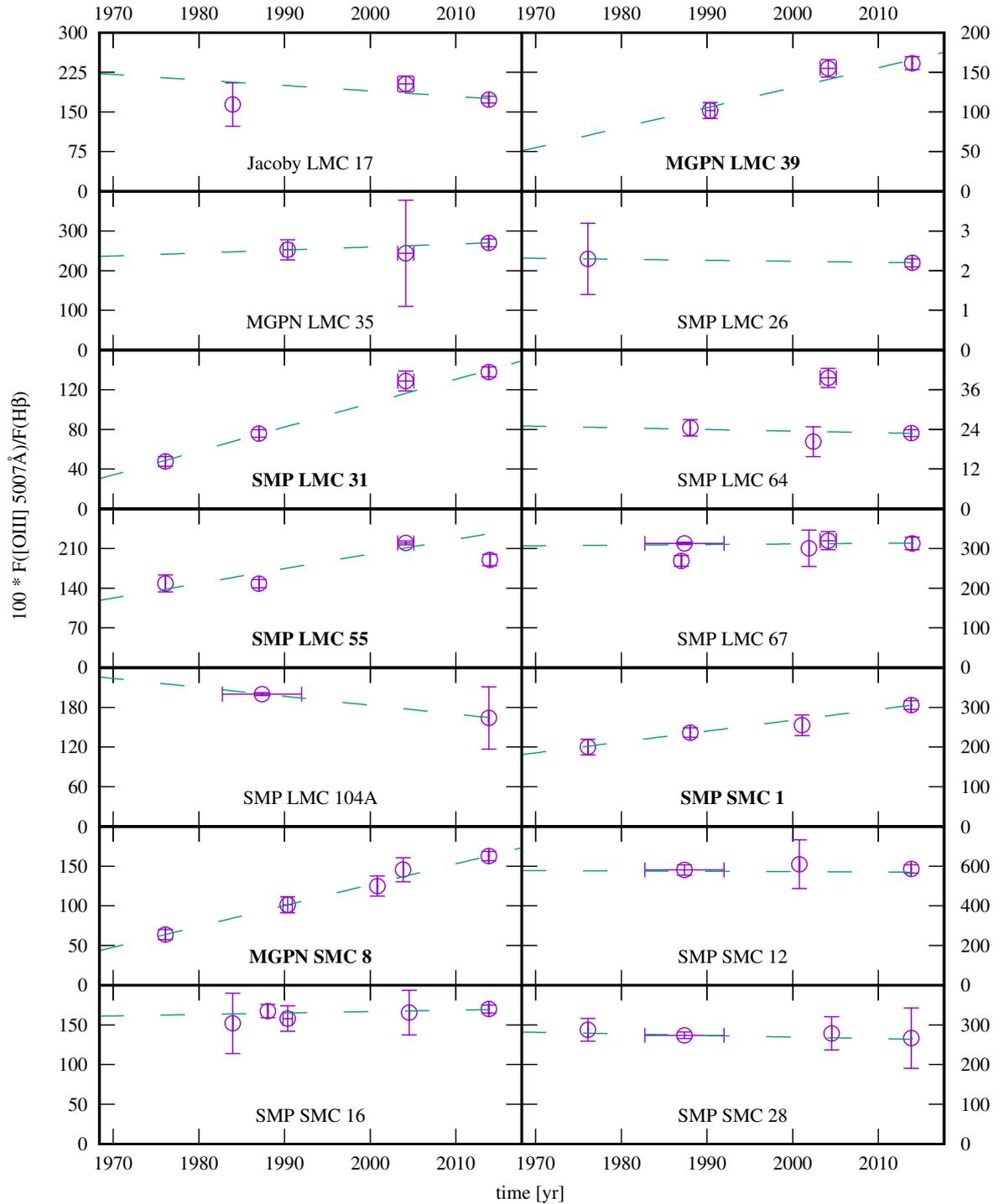}
      \caption{Evolution of [O\,{\sc iii}] 5007\,\AA-to-H$\beta$ line flux ratio in the Magellanic Cloud PNe. The dashed line fits the observed trend. The horizontal error bars correspond to the flux uncertainties. The vertical error bars in the time domain correspond to the observations, of which the dates were not precisely specified. \citet{1991ApJS...75..407M} did not specify dates of their observations precisely, thus the error bars in time domain are large.}
         \label{fig:fluxes}
   \end{figure*}

Hydrogen- and helium-burning post-AGB stars have different evolutionary tracks. Helium-burning tracks show slower evolution by approximately a factor of three in comparison to H-burning tracks for the same remnant mass.

In order to compare the evolution of the central stars of PNe with the evolutionary tracks, one needs to know which shell is active. This can be inferred from the photospheric abundances of a star. A hydrogen-poor atmosphere indicates that the star is a helium burner. 

Most of the H-deficient central stars show [WC] spectra \citep{2011A&A...526A...6W}. They are relatively easy do identify, even if the stellar continuum is unobservable, since they show wide and prominent stellar wind lines imposed on the nebular spectrum \citep{2003A&A...403..659A}. A hydrogen-rich atmosphere does not unambiguously indicate which shell is the main source of the energy in the star. However, we assumed the stars that do not show emission lines to be H burners.

One of the stars in the sample (MGPN\,SMC\,8) is a known [WC] central star \citep{1997A&A...324..674P}. We detected three more [WC] stars in our sample: SMP\,LMC\,31, SMP\,LMC\,55, and SMP\,SMC\,1. They show emission lines of C\,{\sc iii}, C\,{\sc iv}, and He\,{\sc ii}. The He\,{\sc ii} 4686\,\AA\ line in SMP\,LMC\,31 has a full width at half maximum (FWHM) of 13\,\AA, while SMP\,LMC\,55 and SMP\,SMC\,1 have FWHMs of C\,{\sc iv} 4650\,\AA\ lines of 12\,\AA\ and 8\,\AA, respectively. Neither of the new [WC] stars show the C\,{\sc iv} 5806\,\AA\ line. Thus, we classify them as [WC\,11] type \citep{2003A&A...403..659A}.

\section{Heating rates}

We combined new and archival spectra of PNe to study the temperature evolution of their central stars. We used the [O\,{\sc iii}] 5007\,\AA-to-$\mathrm{H}\beta$ line ratio. This ratio is sensitive to the changes of stellar temperature for cool central stars due to a much higher ionization potential of $\mathrm O^+$ than hydrogen. Both lines are relatively strong and located close to each other, so wavelength-dependent calibration errors are reduced.

%{\bf It is also sensitive to the density structure of a PN, however, relative change of this flux is sensitive primarily to the change of the temperature.}

We fit a linear function to the [O\,{\sc iii}] 5007\,\AA- to-$\mathrm{H}\beta$ line flux ratio observed in different epochs for each PN (Fig. \ref{fig:fluxes}). We discarded one point by \citet{2006MNRAS.373..521R} for SMP\,LMC\,64 from the fitting, which exceeded all other points by a factor of two. The fit slope was larger than one, at least by $3 \sigma$ in five PNe, indicating an increasing stellar temperature.

Among the PNe showing the flux evolution, one (MGPN\,SMC\,8) was observed five times, three PNe were observed four times, and one PN has three observations. The observations show a linear increase of the [O\,{\sc iii}] 5007\,\AA-to-$\rm H\beta$ flux ratio in time. An exception is SMP\,LMC\,55. The third observation by \citet{2006MNRAS.373..521R} is significantly higher than the last one. However, the [O\,{\sc iii}] 5007\,\AA-to-H$\beta$ flux ratio determined by \citet{2006MNRAS.373..521R} appears to be systematically higher than fluxes measured by other authors.

We performed spherically symmetric photoionization models of these five PNe with the Cloudy v.\,17.01 code. All the models were constrained by the observed absolute $\mathrm{H}\beta$ flux. We fit the nebular line flux ratios observed in 2013 (Table \ref{tab:fluxes}).

The nebular sizes and stellar luminosities are available for three objects observed with the Hubble Space Telescope by \citet{2003ApJ...597..298V} and \citet{2004ApJ...614..716V}. However, for the PN MGPN\,SMC\,8 temperature and luminosity, determination is most likely affected by the stellar contribution to the He\,{\sc ii} 4686\,\AA\ line flux \citep{2010ApJ...717..562S}. For this and two other PNe, stellar luminosity was a free parameter. The adopted luminosities, or those derived from our photoionization models $L_*,$ are given in Table\,{\ref{tab:tdot}}.

We varied stellar temperature to obtain the best fit. We used atmosphere models by \citet{2003IAUS..210P.A20C}. The grid covers the temperature range of $3.5 - 50$\,kK.

\begin{table*}
 \caption[]{Input parameters and results of the photoionization modeling. Derived stellar temperatures, heating rates, model-dependent luminosities and masses for the central stars, modeled O and N abundances of the PNe, observed luminosities, temperatures, $\mathrm{H}\beta$ fluxes, and diameters.} \label{tab:tdot}
\begin{center}
\begin{tabular}{lccccc}
 \hline \hline
{Name}  &       MGPN\,LMC\,39$^a$               &       SMP\,LMC\,31                    &       SMP\,LMC\,55                 &       MGPN\,SMC\,8                    &       SMP\,SMC\,1                 \\ \hline
 $T_*$ [{kK}]   &       35.7            &       34.8            &       37.3            &       37.3            &       37.2            \\
{$\dot{T_*}$ [$\mathrm{K} \, \mathrm{yr}^{-1}$]}        &        $19.1 \pm 4.7$            &        $18.4 \pm 1.5$                 &        $10.4 \pm 2.1$            &        $24.9 \pm 2.4$                 &        $17.8 \pm 3.4$            \\
{$\log{L_{\mathrm{MOD}}/\mathrm{L}_{\odot}}$}   &        $3.759 \pm 0.014$                 &         $3.791 \pm 0.018$             &        $3.650 \pm 0.074$          &        $3.867 \pm 0.029$              &        $3.784 \pm 0.040$          \\
{${M_{\mathrm{MOD}}/\mathrm{M}_{\odot}}$}       &        $0.570 \pm 0.004$                 &               $0.668 \pm 0.024$       &       $0.613 \pm 0.014$          &       $0.693 \pm 0.016$               &               $0.644 \pm 0.023$      \\
$\log (\mathrm{O}/\mathrm{H})$  &        $-3.50$                &        $-3.92$                 &        $-3.83$                &        $-3.65$                 &        $-3.81$                \\
$\log (\mathrm{N}/\mathrm{H})$          &                       &                        &        $-4.53$                &                       &        $-4.62$                 \\
{$\log{L_*/\mathrm{L}_{\odot}}$}        &        3.50 [1]               &          4.01  [2]     &        4.21   [1]     &         3.42  [1]     &       3.77         [3]     \\
{$\log T_*$ (H\,{\sc i})$^b$ }  &                       &        $28.6 \pm 2.4$    [2]     &                       &        $28.8 \pm 2.6$ [3]             &       $28.8 \pm 2.6$ [3]            \\ 
{$\log F(\mathrm{H}\beta)$}     &       $-13.07$        [4]     &       $-12.91$        [5]     &       $-12.20$        [4]     &       $-13.27$        [6]     &       $-12.85$        [6]     \\
d [arcsec]      &                       &       0.26    [5]     &                       &       0.615   [6]     &       0.15    [6]     \\
\hline
\end{tabular}
\end{center}
\tablebib{[1] this paper, [2] \citet{2003ApJ...597..298V},
[3] \citet{2004ApJ...614..716V},
%[3] \citet{2011AA...530A..90V},
%[4] \citet{2014MNRAS.439.2211K},
[4] \citet{2006MNRAS.373..521R},
[5] \citet{2002ApJ...575..178S},
[6] \citet{2003ApJ...596..997S},
$^a$central star without emission lines
$^b$computed using black-body model
}
\end{table*}

We used nebular abundances by \citet{2006AA...456..451L} as an input parameter, but varied it to better fit the observed emission line fluxes. For some cases (e.g., MGPN\,LMC\,39, SMP\,LMC\,31, and SMP\,SMC\,1) we were not able to achieve the observed [O\,{\sc iii}] 5007\,\AA-to-$\mathrm{H}\beta$ line flux ratio with the O/H abundances derived by \citet{2006AA...456..451L}. At low metallicities, the [O\,{\sc iii}] 5007\,\AA-to-H$\beta$ line flux ratio depends on the nebular O/H abundance and stellar temperature \citep{2007arXiv0704.0348S}. For the average SMC oxygen abundance, the [O\,{\sc iii}] 5007\,\AA-to-H$\beta$ flux ratio reaches a maximum of five at a temperature of about 100\,kK \citep{2003ApJ...596..997S}. For the LMC, the models reach maximum of ten at similar stellar temperatures. 

Table\,\ref{tab:tdot} shows the input parameters and results of the photoionization modeling of the five objects. The modeled line fluxes are compared to the fluxes measured in 2013 in Table \ref{tab:models}.

We varied stellar temperature to fit the [O\,{\sc iii}] 5007\,\AA-to-$\mathrm{H}\beta$ line flux ratio observed in other epochs. After we obtained the temperatures in different years, we used them to compute heating rates $\dot{T}_*$. The uncertainty of the measured [O\,{\sc iii}] 5007\,\AA-to-$\mathrm{H}\beta$ line flux ratio propagates to the measured temperatures in different epochs and subsequently to the heating rates presented in Table \ref{tab:tdot}.

We computed heating rates from the evolutionary tracks as a function of stellar temperature. We used models by \citet{2016A&A...588A..25M} for the star that does not show emission lines, and \citet{1994ApJS...92..125V} for four [WC] central stars. \citet{2016A&A...588A..25M} computed only three He-burning models spanning a very low range of masses.

Heating rates derived from the evolutionary tracks were interpolated to fit observed heating rates. We determined model-dependent luminosities $L_{\mathrm{MOD}}$ and corresponding model-dependent masses $M_{\mathrm{MOD}}$. The errors of $L_{\mathrm{MOD}}$ and $M_{\mathrm{MOD}}$ shown in Table\,{\ref{tab:tdot}} result from the uncertainties in $\dot{T}_*$.

We did not include systematic errors in measured heating rates in Table\,\ref{tab:tdot}. These depend mostly on the determination of the O/H abundance. For an adopted uncertainty of the log\,(O/H) of 0.3 in dex, the model-dependent stellar mass $M_{\mathrm{MOD}}$ is accurate to about $0.04 \, \mathrm{M}_{\odot}$, and log of luminosity to 0.15 dex. Moreover, the model-dependent masses for [WC] stars derived using tracks by \citet{1994ApJS...92..125V} may be overestimated by $0.038\,M_{\odot}$ \citep{2014A&A...566A..48G}.

Heating rates derived from the evolutionary tracks do not vary strongly with stellar temperature. The uncertainty of stellar temperature determination affects the derived model-dependent parameters to a much lower extent than the uncertainty in the determination of O/H abundance.

\section{Discussion}

   \begin{figure}
   \centering
   \includegraphics[width=\hsize]{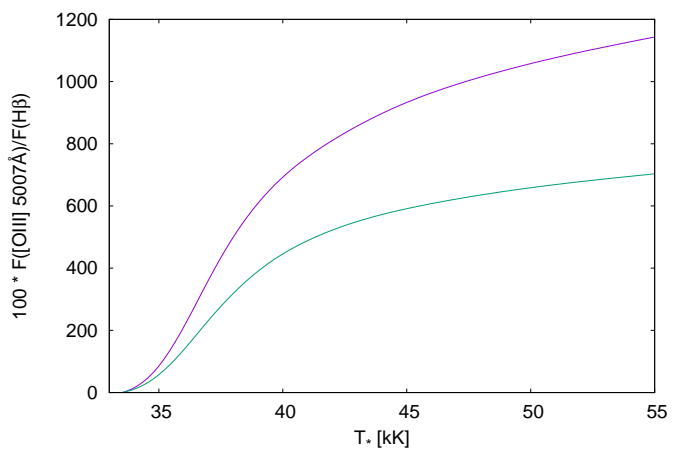}
      \caption{Two models of the evolution of the [O\,{\sc iii}] 5007\,\AA-to-$\rm H\beta$ flux ratio with the temperature of the central star in a PN. The model parameters are identical to MGPN\,LMC\,39, and the second model the (O/H) abundance is lowered by 0.3 in dex. 
              }
         \label{fig:o3evol}
   \end{figure}

We attempted to select PNe with relatively cool central stars to study their evolution. Our sample fulfils the criterion of [O\,{\sc iii}] 5007\,\AA- to-$\mathrm H \beta$ ratio ${\lesssim}3$ for all but one object. However, this criterion appeared to be insufficient in some cases. The [O\,{\sc iii}] 5007\,\AA\ line may be weak in hot PNe, due to photoionization of $\mathrm{O}^{++}$ to $\mathrm{O}^{+++}$ \citep{2001A&A...378..958M,1989A&A...213..274S}. Also, the [O\,{\sc iii}] 5007\,\AA\ line can be weak in PNe with low O abundances, despite a high temperature. This appears to be the case for three PNe in our sample: Jacoby\,LMC\,17, MGPN\,LMC\,35, and SMP\,SMC\,28, which show prominent He\,{\sc ii} 4686\,\AA\ lines. These PNe are not expected to show rapid evolution of the [O\,{\sc iii}] 5007\,\AA\ flux. 

Young PNe can mimic symbiotic stars due to high nebular density and strong dust emission. \citet{2017A&A...606A.110I} listed two PNe from our sample, SMP\,LMC\,31 and SMP\,LMC\,104A, as symbiotic star candidates. However, we did not find any indication of the cool giant in their spectra. In addition, SMP\,LMC\,31 show stellar wind lines consistent with a [WC] star.

Eleven out of fourteen PNe in our sample may be considered as young PNe. Five of them show an evolution of the [O\,{\sc iii}] 5007\,\AA\ line flux. The five central stars showing temperature evolution are most likely the most massive in our sample, and thus the most luminous. An independent luminosity determination confirms that these stars show the highest luminosities (Table \ref{tab:lum}). 

The remaining six central stars may evolve too slowly to show significant evolution of the [O\,{\sc iii}] 5007\,\AA- to-H$\beta$ flux ratio on a timescale of decades. The collected observations are sensitive down to heating rates of about $5 - 10 \, \mathrm{K} \, \mathrm{yr}^{-1}$. Lower heating rates, corresponding to the initial masses below $1.25 \, \mathrm M_{\odot}$ (final masses below $0.532 \, \mathrm M_{\odot}$) \citep{2019IAUS..343...36M}, would not be detected.

Figure \ref{fig:o3evol} shows the modeled evolution of nebular [O\,{\sc iii}] 5007\,\AA- to-$\rm H\beta$ flux ratio with increasing stellar temperature for two different O/H abundance ratios. The [O\,{\sc iii}] 5007\,\AA-to-$\rm H\beta$ flux ratio scales approximately with the O abundance. The flux ratio does not increase linearly throughout the whole temperature range of the central star, but it can be approximated as linear in sufficiently small temperature intervals. The heating rate is highest between 35 and 38\,kK. Then it slows down. The five central stars of which the nebulae show flux evolution have stellar temperatures in this range.

The central star of MGPN\,LMC\,39 is probably the most massive H-burning central star in our sample. Its final mass of $0.57 \, \mathrm{M}_{\odot}$ is the same as the peak of the mass distribution for the central stars of Galactic PNe. The remaining H-burning central stars in our sample are less massive. This suggests that central stars in the Magellanic Clouds are less massive than the stars in the Galaxy. \citet{2014A&A...567A..15H} derived heating rates of $45 \pm 7 \, \mathrm{K} \, \mathrm{yr}^{-1}$ for Galactic PN Hen\,2-260, significantly higher than for Magellanic Cloud central stars.

We compared the model-dependent luminosities $L_{\mathrm{MOD}}$ with luminosities determined independently using different methods: using Zanstra temperatures and a black-body fit ($\log T_*$ (H\,{\sc i})), by fitting spectral energy distribution (SED) ($L_{\mathrm{SED}}$) and by photoionization modeling ($L_*$) (Tab.\,\ref{tab:lum}). Unfortunately, the luminosities yielded from observations using different methods show a relatively large scatter.

The model-dependent luminosity agrees with the observed luminosity for SMP\,SMC\,1 and is within the observed range of stellar luminosities for other objects. For SMP\,LMC\,55, the model-dependent luminosity is close to the luminosity derived from photoionization modeling by \citet{1991ApJ...367..115D}, but lower by 0.56 dex from the luminosity of our photoionization model. However, the absolute $\mathrm{H}\beta$ flux used by \citet{1991ApJ...367..115D} was 0.46 lower in log than that used by us from \citet{2006MNRAS.373..521R}. If this discrepancy were real, this would indicate an unexpected change in luminosity of this object.

We are not able to discriminate between \citet{2016A&A...588A..25M} and \citet{1994ApJS...92..125V} or \citet{1995A&A...299..755B}.  \citet{2016A&A...588A..25M} models were faster and more luminous for the same masses. However, these models all show similar evolutionary speeds for the same luminosity of the central star (but different masses). We would be able to discriminate between these models if we obtained an independent mass determination.

\begin{table*}
 \caption[]{Luminosities of the central stars of PNe obtained using different methods: model-dependent luminosities determined from heating rates, luminisities derived from Zanstra temperatures and stellar magnitudes, from the  integrated spectral energy distribution, and constrained from abolute $\mathrm{H\beta}$ flux and nebular modeling. For SMP\,LMC\,64 and SMP\,LMC\,104A, no data were found in the literature.}\label{tab:lum}
\begin{center}
\begin{tabular}{lcccccc}
 \hline \hline
        &  $\log L_\mathrm{MOD}$     &  {$\log L_*/\mathrm{L}_{\odot}$ (H\,{\sc i})}    &       {$\log L_{\mathrm{SED}} /\mathrm{L}_{\odot}$}   &       {$\log L_{\mathrm{SED}} /\mathrm{L}_{\odot}$}  &       {$\log L_* /\mathrm{L}_{\odot}$}        & {$\log L_* /\mathrm{L}_{\odot}$} \\
ref. & [1] & [2] & [3] & [4] & [5] & [1] \\
\hline
Jacoby LMC 17     &      &              &       3.39    &               &       \\
MGPN LMC 35     &               &    &  3.20    &               &       & \\
{\bf MGPN LMC 39}       &  {\bf 3.759}  &               &       {\bf 3.70}      &       {\bf 3.47}   &       & {\bf 3.50} \\
SMP LMC 26      &   &           &       3.60    &       3.32    &       & \\
{\bf SMP LMC 31}    & {\bf 3.791}       &       {\bf 4.01}      &       {\bf 3.90}   &       {\bf 3.64}      &       & \\
{\bf SMP LMC 55}        &  {\bf 3.650} &                &               &               &       {\bf 3.69} & {\bf 4.21} \\
%SMP LMC 64     &               &               &               &               \\
SMP LMC 67      &   &           &               &       2.86    &       3.33 & \\
%SMP LMC 104A   &               &               &               &               \\
{\bf MGPN SMC 8}        &  {\bf 3.867} &        {\bf 4.33}      &               &       {\bf 3.49}   &       & {\bf 3.42}    \\
{\bf SMP SMC 1} &   {\bf 3.784} &       {\bf 3.77}      &               &               &       &         \\
SMP SMC 12      &   &   2.84    &               &               &       &         \\
SMP SMC 16      &   &           &               &       3.20    &       &         \\
SMP SMC 28      &   &           &               &               &       3.28 &       \\
\hline
\end{tabular}
\tablebib{[1] this paper, [2] \citet{2003ApJ...597..298V,2004ApJ...614..716V},
[3] \citet{2011AA...530A..90V} ,
[4] \citet{2014MNRAS.439.2211K},
[5] \citet{1991ApJ...367..115D}}
\end{center}
\end{table*}

The [WC] central stars constitute about 15\% of Galactic central stars of PNe. This ratio was suggested to be as low as 5\% in the Magellanic Clouds \citep{1997A&A...324..674P}. \citet{1997A&A...324..674P} reported only two [WC] central stars in the Small Magellanic Cloud (SMC) and six in the Large Magellanic Cloud (LMC). Five stars are intermediate types [WC\,4] or [WC\,4-5], and only one is [WC\,9] \citep{1988MNRAS.234..583M}. \citet{2020ApJ...888...54M} and \citet{2020arXiv200604333M} announced the discovery of two more [WC\,11] stars in the LMC. They suggested that a large population of late-type [WC] stars may exist, undetected due to observational selection. Late-type [WC] stars show faint emission lines, which could be missed in observations performed with smaller telescopes. We discovered three new late-type [WC] stars. 

Interestingly, the four [WC] PNe observed by us show rapid evolution. This implies relatively high stellar mass. 
\citet{2015A&A...573A..65H} observed a relatively rapid change of the [O\,{\sc iii}] 5007\,\AA-to-$\mathrm{H}\beta$ flux ratio for the Galactic PNe M\,1-11 and M\,1-12 containing [WC] central stars. This suggests that the evolution of Galactic [WC] stars also is faster than H-rich central stars and originate from more massive progenitors. This was independently confirmed by \citet{2011A&A...526A...6W}, who found that H-poor central stars are more concentrated toward the Galactic center and Galactic plane than the H-rich group. The DB white dwarfs, which are the possible progeny of [WC] central stars, show an average mass of $0.651 \, \mathrm{M}_{\odot}$, higher than the $0.598 \, \mathrm{M}_{\odot}$ derived for DA white dwarfs, which originate from H-rich central stars \citep{2013ApJS..204....5K}.

\begin{acknowledgements}
MH thanks the Ministry of Science and Higher Education (MSHE) of the Republic of Poland for granting funds for the Polish contribution to the International LOFAR Telescope (MSHE decision no. DIR/WK/2016/2017/05-1) and for maintenance of the LOFAR PL-612 Baldy (MSHE decision no.~59/E-383/SPUB/SP/2019.1). Polish participation in SALT is funded by the MSHE grant No. DIR/WK/2016/07. MH acknowledges financial support from National Science Centre, Poland, grant No. 2016/23/B/ST9/01653. Some of the observations reported in this paper were obtained with the Southern African Large Telescope (SALT) under programme 2013-2-POL\_OTH-001 (PI M. Hajduk).  
\end{acknowledgements}

\bibliographystyle{aa} % style aa.bst
\bibliography{references} % your references Yourfile.bib

\begin{appendix}
%\chapter{Line fluxes}

\section{Line fluxes}
%\renewcommand{\thetable}{\Alph{chapter}.\arabic{table}}
%\counterwithin{table}{section}
\numberwithin{table}{section}

\clearpage
\onecolumn
\begin{landscape}
\begin{longtable}{*{17}{c}}
\caption{The measured and dereddened line fluxes in the LMC PNe with the relevant uncertainties. The value of the $E(B-V)$ color excess applied for dereddening the fluxes is given for each PN.}\label{tab:fluxes} \\
\hline \hline
&               &               \multicolumn{3}{c}{Jacoby\,LMC\,17}     &               \multicolumn{3}{c}{MGPN\,LMC\,35}       &                       \multicolumn{3}{c}{MGPN\,LMC\,39}       &                       \multicolumn{3}{c}{SMP\,LMC\,26}        &       \multicolumn{3}{c}{SMP\,LMC\,31} \\
        &       {$\lambda$}     &       {$F(\lambda)$}  &       {$I(\lambda)$}  &       {$\delta I$}     &       {$F(\lambda)$}  &       {$I(\lambda)$}  &       {$\delta I$}     &       {$F(\lambda)$}  &       {$I(\lambda)$}  &       {$\delta I$}     &       {$F(\lambda)$}  &       {$I(\lambda)$}  &       {$\delta I$}     &       {$F(\lambda)$}  &       {$I(\lambda)$}  &       {$\delta I$}     
\\ 
\hline
$E(B-V)$        &   &           &       0.36    &               &               &       0.10    &               &               &       0.45    &               &               &       0.19    &               &               &       0.37    &                               \\
$\rm H\gamma$   &       4343    &       38.6    &       45.5    &       1.7     &       43.7    &       45.8    &       3.4     &               &               &               &               &               &               &       43.2    &       51.1    &       1.9     \\
{[}O\,{\sc iii}{]}      &       4363    &       20.6    &       24.1    &       1.0     &       30.1    &       31.5    &       3.3     &               &               &               &               &               &               &       0.9     &       1.1     &       0.2     \\
He\,{\sc i}     &       4388    &               &               &               &               &               &               &               &               &               &               &               &               &       0.4     &       0.5     &       0.1     \\
He\,{\sc i}     &       4471    &               &               &               &               &               &               &       2.6     &       3.1     &       0.4     &               &               &       &       2.8     &       3.2     &       0.2     \\
He\,{\sc ii}    &       4541    &       2.0     &       2.2     &       0.3     &               &               &               &               &               &               &               &               &               &       1.5     &       1.6     &       0.2     \\
C\,{\sc iv}     &       4650    &               &               &               &               &               &               &       1.4$^a$ &       1.6$^a$ &       0.3$^a$ &               &               &               &       1.5$^a$ &       1.6$^a$ &       0.2$^a$ \\
He\,{\sc ii}    &       4686    &       10.2    &       10.8    &       0.6     &       58.3    &       59.2    &       2.7     &               &               &               &               &               &               &               &               &               \\
{[}Ar\,{\sc iv}{]}/He{\sc i}    &       4712    &       1.0     &       1.1     &       0.2     &               &               &               &               &               &               &               &               &               &       0.5     &       0.5     &       0.1     \\
{[}Ne\,{\sc iv}{]}      &       4726    &       1.4     &       1.4     &       0.3     &               &               &               &               &               &               &               &               &               &               &               &               \\
{[}Ar\,{\sc iv}{]}      &       4741    &       0.7     &       0.7     &       0.2     &               &               &               &               &               &               &               &               &               &               &               &               \\
$\rm H\beta$    &       4861    &       100.0   &       100.0   &       3.6     &       100.0   &       100.0   &       4.0     &       100.0   &       100.0   &       5.0     &       100.0   &       100.0   &       5.0     &       100.0   &       100.0   &       3.6     \\
He\,{\sc i}     &       4922    &               &               &               &               &               &               &               &               &               &               &               &               &0.9    &       0.9     &       0.1     \\
{[}O\,{\sc iii}{]}      &       4931    &               &               &               &               &               &               &               &               &               &               &               &               &               &               &               \\
{[}O\,{\sc iii}{]}      &       4959    &       65.1    &       63.0    &       2.3     &       88.5    &       87.7    &       3.5     &       54.7    &       52.5    &       2.7     &       0.5     &       0.5     &       0.1     &       48.7    &       47.1    &       1.7     \\
{[}O\,{\sc iii}{]}      &       5007    &       182.0   &       173.4   &       6.2     &       274.2   &       270.5   &       9.9     &       170.0   &       160.1   &       8.0     &       2.3     &       2.2     &       0.1     &       145.6   &       138.6   &       5.0     \\
{[}N\,{\sc i}{]}        &       5197    &               &               &               &               &               &               &               &               &               &       0.6     &       0.6     &       0.1     &0.3    &       0.3     &       0.1     \\
He\,{\sc ii}    &       5412    &       6.2     &       5.2     &       0.2     &       6.4     &       6.1     &       0.8     &               &               &               &               &               &               &               &               &               \\
{[}Cl\,{\sc iii}{]}     &       5537    &               &               &               &               &               &               &               &               &               &               &               &               &       0.1     &       0.1     &       0.1     \\
C\,{\sc iii}    &       5696    &               &               &               &                       &               &               &               &               &               &               &               &               &       0.8$^a$ &       0.6$^a$ &       0.1$^a$ \\
{[}N\,{\sc ii}{]}       &       5755    &       4.1     &       3.2     &       0.2     &               &               &               &       2.2     &       1.6     &       0.2     &       2.7     &       2.4     &       0.1     &3.6    &       2.8     &       0.2     \\
He\,{\sc i}     &       5876    &       5.3     &       4.0     &       0.2     &       10.9    &       10.1    &       1.0     &       12.4    &       8.7     &       0.5     &       1.8     &       1.5     &       0.1     &       13.7    &       10.3    &       0.4     \\
{[}O\,{\sc i}{]}        &       6302    &       13.2    &       9.1     &       0.3     &       5.2     &       4.7     &       0.5     &       2.6     &       1.6     &       0.2     &       2.3     &       1.9     &       0.1     &       3.3     &       2.2     &       0.1     \\
{[}S\,{\sc iii}{]}      &       6312    &       2.7     &       1.9     &       0.1     &               &               &               &       2.1     &       1.3     &       0.1     &               &               &               &       1.0     &       0.7     &       0.1     \\
{[}O\,{\sc i}{]}        &       6347    &               &               &               &               &               &               &               &               &               &               &               &               &       0.3     &       0.2     &       0.1     \\
{[}O\,{\sc i}{]}        &       6365    &       4.7     &       3.2     &       0.1     &       3.9     &       3.5     &       0.7     &       0.7     &       0.4     &       0.1     &       0.5     &       0.5     &       0.1     &       1.2     &       0.8     &       0.1     \\
{[}N\,{\sc ii}{]}       &       6548    &       15.2    &       10.0    &       0.4     &       14.1    &       12.5    &       0.7     &       15.7    &       9.3     &       0.5     &       41.6    &       33.6    &       1.7     &       23.6    &       15.5    &       0.6     \\
$\rm H\alpha$   &       6563    &       433.5   &       286.0   &       10.1    &       321.9   &       286.0   &       10.2    &       482.6   &       286.0   &       14.2    &       355.0   &       286.0   &       14.2    &       438.5   &       286.0   &       10.2    \\
{[}N\,{\sc ii}{]}       &       6584    &       34.5    &       22.7    &       0.8     &       28.3    &       25.1    &       1.1     &       42.9    &       25.3    &       1.3     &       119.0   &       95.7    &       4.7     &       71.7    &       46.6    &       1.7     \\
He\,{\sc i}     &       6678    &       1.7     &       1.1     &       0.1     &       2.9     &       2.6     &       0.5     &       3.9     &       2.3     &       0.2     &       0.6     &       0.5     &       0.9     &       4.3     &       2.8     &       0.2     \\
{[}S\,{\sc ii}{]}       &       6716    &       2.0     &       1.3     &       0.1     &       4.8     &       4.2     &       0.6     &       3.9     &       0.6     &       0.1     &       0.4     &       0.3     &       0.1     &       0.5     &       0.3     &       0.1     \\
{[}S\,{\sc ii}{]}       &       6730    &       2.4     &       1.5     &       0.1     &       4.5     &       4.0     &       0.5     &               &               &               &       1.7     &       1.4     &       0.1     &       1.1     &       0.7     &       0.1     \\
He\,{\sc i}     &       7002    &               &               &               &               &               &               &               &               &               &               &               &               &               &               &               \\
{[}Ar\,{\sc v}{]}       &       7007    &       3.1     &       1.9     &       0.1     &       4.0     &       3.5     &       0.5     &       11.4    &       6.1     &       0.4     &       0.4     &       0.3     &       0.1     &               &               &               \\
He\,{\sc i}     &       7065    &       3.3     &       2.0     &       0.1     &       7.3     &       6.4     &       0.6     &               &               &               &       0.8     &       0.6     &       0.1     &               &               &               \\
{[}Ar\,{\sc iii}{]}     &       7100    &               &               &               &               &               &               &       11.2    &       5.9     &       0.3     &               &               &               &       0.5     &       0.3     &       0.1     \\
{[}Ar\,{\sc iii}{]}     &       7136    &       7.2     &       4.3     &       0.2     &       7.4     &       6.4     &       0.7     &               &               &               &       1.0     &       0.8     &       0.1     &       9.2     &       5.4     &       0.2     \\
{[}Ar\,{\sc iv}{]}      &       7172    &       1.5     &       0.9     &       0.1     &               &               &               &       2.4     &       1.3     &       0.1     &               &               &               &               &               &               \\
{[}Ar\,{\sc iv}{]}      &       7237    &               &               &               &               &               &               &               &               &               &               &               &               &       2.0     &       1.2     &       0.1     \\
He\,{\sc i}     &       7281    &               &               &               &               &               &               &       1.0     &       0.5     &       0.1     &               &               &               &       1.1     &       0.6     &       0.1     \\
{[}O\,{\sc ii}{]}       &       7320    &       8.5     &       5.0     &       0.2     &               &               &               &       65.9    &       33.5    &       1.7     &       23.4    &       17.7    &       0.9     &       46.2    &       26.6    &       1.0     \\
{[}O\,{\sc ii}{]}       &       7330    &       6.9     &       4.0     &       0.2     &               &               &               &       49.0    &       24.9    &       1.3     &       17.6    &       13.3    &       0.7     &       36.8    &       21.1    &       0.8     \\
\hline
$^a$ stellar line
\end{longtable}

\setcounter{table}{0} 
\newpage
\begin{longtable}{*{17}{c}}
\caption{The measured and dereddened line fluxes in the MC PNe with the relevant uncertainties. The value of the $E(B-V)$ color excess applied for dereddening the fluxes is given for each PN.} \\
\hline \hline
        &               &       \multicolumn{3}{c}{SMP\,LMC\,55}                        &                       \multicolumn{3}{c}{SMP\,LMC\,64}                &               \multicolumn{3}{c}{SMP\,LMC\,67}                &               \multicolumn{3}{c}{SMP\,LMC\,104A}              &               \multicolumn{3}{c}{MGPN\,SMC\,8}                \\
        &       {$\lambda$}     &       {$F(\lambda)$}  &       {$I(\lambda)$}  &       {$\delta I$}     &               {$F(\lambda)$}  &       {$I(\lambda)$}  &       {$\delta I$}     &       {$F(\lambda)$}  &       {$I(\lambda)$}  &       {$\delta I$}     &       {$F(\lambda)$}  &       {$I(\lambda)$}  &       {$\delta I$}     &       {$F(\lambda)$}  &       {$I(\lambda)$}  &       {$\delta I$}     \\
\hline
\hline
$E(B-V)$   &            &               &       0.25    &               &                       &       0.52    &               &               &       0.22    &               &               &       0.23    &               &               &       0.26                    \\
$\rm H\gamma$   &       4343    &               &               &               &               56.5    &       71.8    &       1.9     &               &               &               &       51.0    &       56.7    &       16.1    &       35.1    &       39.5    &       1.5     \\
{[}O\,{\sc iii}{]}      &       4363    &       1.3     &       1.5     &       0.2     &               3.2     &       4.1     &       0.2     &       1.4     &       1.6     &       0.2     &       21.1    &       23.2    &       6.6     &       2.3     &       2.6     &       0.3     \\
He\,{\sc i}     &       4388    &       0.5     &       0.6     &       0.1     &               1.0     &       1.3     &       0.2     &       0.6     &       0.6     &       0.1     &       0.2     &       0.2     &       0.1     &               &               &               \\
He\,{\sc i}     &       4471    &       3.4     &       3.7     &       0.2     &               4.4     &       5.3     &       0.2     &       4.3     &       4.5     &       0.3     &       1.5     &       1.5     &       0.5     &       4.4     &       4.8     &       0.3     \\
He\,{\sc ii}    &       4541    &       0.4     &       0.4     &       0.1     &               1.4     &       1.6     &       0.1     &               &               &               &       2.2     &       2.3     &       0.7     &       1.9     &       2.0     &       0.2     \\
C\,{\sc iv}     &       4650    &       1.1$^a$ &       1.1$^a$ &       0.1$^a$ &                       &               &               &               &               &               &               &               &               &       119.5$^a$       &       125.6$^a$       &       4.5$^a$ \\
He\,{\sc ii}    &       4686    &       1.1$^a$ &       1.1$^a$ &       0.1$^a$ &                       &               &               &               &               &               &       62.2    &       65.0    &       18.4    &       24.9$^a$        &       26.0$^a$        &       1.0$^a$ \\
{[}Ar\,{\sc iv}{]}/He{\sc i}    &       4712    &       0.5     &       0.6     &       0.1     &               0.5     &       0.6     &       0.1     &       0.6     &       0.6     &       0.1     &       1.5     &       1.5     &       0.5     &               &               &               \\
{[}Ne\,{\sc iv}{]}      &       4726    &               &               &               &                       &               &               &               &               &               &       1.5     &       1.5     &       0.5     &               &               &               \\
{[}Ar\,{\sc iv}{]}      &       4741    &               &               &               &                       &               &               &               &               &               &       1.1     &       1.1     &       0.4     &               &               &               \\
$\rm H\beta$    &       4861    &       100.0   &       100.0   &       5.0     &               100.0   &       100.0   &       2.6     &       100.0   &       100.0   &       5.0     &       100.0   &       100.0   &       28.3    &       100.0   &       100.0   &       3.6     \\
He\,{\sc i}     &       4922    &       0.9     &       0.8     &       0.1     &               2.4     &       2.3     &       0.1     &       1.3     &       1.3     &       0.1     &       0.6     &       0.6     &       0.2     &       1.0     &       1.0     &       0.2     \\
{[}O\,{\sc iii}{]}      &       4931    &               &               &               &                       &               &               &               &               &               &       0.5     &       0.5     &       0.2     &               &               &               \\
{[}O\,{\sc iii}{]}      &       4959    &       64.4    &       62.9    &       3.2     &               6.3     &       6.1     &       0.2     &       105.1   &       103.4   &       5.2     &       56.8    &       54.4    &       15.4    &       53.7    &       52.5    &       1.9     \\
{[}O\,{\sc iii}{]}      &       5007    &       193.8   &       187.3   &       9.3     &               22.8    &       21.3    &       0.6     &       320.4   &       303.7   &       15.1    &       172.1   &       163.8   &       46.4    &       168.2   &       162.5   &       5.8     \\
{[}N\,{\sc i}{]}        &       5197    &       0.5     &       0.4     &       0.1     &               1.3     &       1.1     &       0.1     &       1.4     &       1.4     &       0.1     &               &               &               &               &               &               \\
He\,{\sc ii}    &       5412    &               &               &               &                       &               &               &               &               &               &               &               &               &               &               &               \\
{[}Cl\,{\sc iii}{]}     &       5537    &       0.2     &       0.2     &       0.1     &                       &               &               &       0.3     &       0.2     &       0.1     &               &               &               &               &               &               \\
C\,{\sc iii}    &       5696    &       0.4$^a$ &       0.3$^a$ &       0.1$^a$ &                       &               &               &               &               &               &               &               &               &       51.1$^a$        &       43.0$^a$        &       1.5$^a$ \\
{[}N\,{\sc ii}{]}       &       5755    &       2.3     &       1.9     &       0.1     &                       &               &               &       4.1     &       3.5     &       0.2     &       0.7     &       0.6     &       0.2     &               &               &               \\
He\,{\sc i}     &       5876    &       13.3    &       10.9    &       0.6     &                       &               &               &       16.0    &       12.9    &       0.7     &       5.0     &       4.2     &       1.2     &               &               &               \\
{[}O\,{\sc i}{]}        &       6302    &       2.8     &       2.2     &       0.2     &               6.4     &       3.8     &       0.1     &       2.1     &       1.7     &       0.1     &       2.9     &       2.3     &       0.7     &       6.8     &       5.2     &       0.2     \\
{[}S\,{\sc iii}{]}      &       6312    &       1.2     &       0.9     &       0.1     &               5.1     &       3.0     &       0.1     &       0.7     &       0.6     &       0.1     &       1.6     &       1.2     &       0.4     &       1.1     &       0.9     &       0.1     \\
{[}O\,{\sc i}{]}        &       6347    &               &               &               &                       &               &               &               &               &               &               &               &               &               &               &               \\
{[}O\,{\sc i}{]}        &       6365    &       0.9     &       0.7     &       0.1     &               2.1     &       1.2     &       0.1     &       0.7     &       0.6     &       0.1     &       0.9     &       0.7     &       0.3     &       2.3     &       1.7     &       0.1     \\
{[}N\,{\sc ii}{]}       &       6548    &       34.0    &       25.4    &       1.3     &               37.4    &       20.5    &       0.6     &       102.3   &       80.4    &       4.0     &       5.3     &       4.1     &       1.2     &       37.1    &       27.6    &       1.0     \\
$\rm H\alpha$   &       6563    &       384.3   &       286.0   &       14.2    &               522.8   &       286.0   &       7.2     &       364.8   &       286.0   &       14.2    &       372.3   &       286.0   &       80.9    &       385.8   &       286.0   &       10.2    \\
{[}N\,{\sc ii}{]}       &       6584    &       102.3   &       75.9    &       3.8     &               50.3    &       27.3    &       0.7     &       309.6   &       240.3   &       11.9    &       3.0     &       2.3     &       0.7     &       116.3   &       86.0    &       3.1     \\
He\,{\sc i}     &       6678    &       4.0     &       2.9     &       0.2     &               2.7     &       1.4     &       0.1     &       4.7     &       3.7     &       0.2     &       0.9     &       0.7     &       0.2     &       0.9     &       0.7     &       0.1     \\
{[}S\,{\sc ii}{]}       &       6716    &       1.7     &       1.2     &       0.1     &               0.6     &       0.3     &       0.1     &       3.4     &       2.6     &       0.2     &       0.3     &       0.2     &       0.1     &       10.7    &       7.8     &       0.3     \\
{[}S\,{\sc ii}{]}       &       6730    &       3.4     &       2.4     &       0.2     &               0.9     &       0.5     &       0.1     &       5.0     &       3.8     &       0.2     &       0.5     &       0.3     &       0.1     &       14.4    &       10.5    &       0.4     \\
He\,{\sc i}     &       7002    &       0.2     &       0.1     &       0.3     &               0.5     &       0.2     &       0.1     &       0.1     &       0.1     &       0.1     &               &               &               &               &               &               \\
{[}Ar\,{\sc v}{]}       &       7007    &               &               &               &                       &               &               &               &               &               &       2.3     &       1.7     &       0.5     &               &               &               \\
He\,{\sc i}     &       7065    &       9.9     &       6.9     &       2.3     &               4.6     &       2.2     &       0.1     &       6.2     &       4.6     &       0.3     &       3.8     &       2.8     &       0.8     &       6.7     &       4.7     &       0.2     \\
{[}Ar\,{\sc iii}{]}     &       7100    &               &               &               &                       &               &               &               &               &               &               &               &               &               &               &               \\
{[}Ar\,{\sc iii}{]}     &       7136    &       28.7    &       6.3     &       2.1     &               5.7     &       2.7     &       0.1     &       9.2     &       6.7     &       0.4     &       3.1     &       2.2     &       0.7     &       6.6     &       4.6     &       0.2     \\
{[}Ar\,{\sc iv}{]}      &       7172    &       21.8    &       0.7     &       0.5     &               0.4     &       0.2     &       0.1     &               &               &               &       0.6     &       0.5     &       0.2     &               &               &               \\
{[}Ar\,{\sc iv}{]}      &       7237    &               &               &               &                       &               &               &               &               &               &               &               &               &               &               &               \\
He\,{\sc i}     &       7281    &       1.0     &       0.7     &       0.5     &               0.7     &       0.3     &       0.1     &       0.9     &       0.7     &       0.1     &       0.3     &       0.2     &       0.1     &               &               &               \\
{[}O\,{\sc ii}{]}       &       7320    &       28.7    &       19.6    &       6.0     &               24.5    &       11.2    &       0.3     &       7.1     &       5.2     &       0.3     &       3.8     &       2.8     &       0.8     &       16.0    &       10.8    &       0.4     \\
{[}O\,{\sc ii}{]}       &       7330    &       21.8    &       14.9    &       4.8     &               19.2    &       8.8     &       0.3     &       6.0     &       4.4     &       0.3     &       3.1     &       2.2     &       0.7     &       12.0    &       8.2     &       0.3     \\
\hline
$^a$ stellar line
\end{longtable}

\setcounter{table}{0} 
\newpage
\begin{longtable}{*{14}{c}}
\caption[]{The measured and dereddened line fluxes in the SMC PNe with the relevant uncertainties. The value of the $E(B-V)$ color excess applied for dereddening the fluxes is given for each PN.} \\
\hline \hline
        &               &               \multicolumn{3}{c}{SMP\,SMC\,1}                 &               \multicolumn{3}{c}{SMP\,SMC\,12}        &       \multicolumn{3}{c}{SMP\,SMC\,16}        &       \multicolumn{3}{c}{SMP\,SMC\,28}                \\
        &       {$\lambda$}     &       {$F(\lambda)$}  &       {$I(\lambda)$}  &       {$\delta I$}     &       {$F(\lambda)$}  &       {$I(\lambda)$}  &       {$\delta I$}     &       {$F(\lambda)$}  &       {$I(\lambda)$}  &       {$\delta I$}     &       {$F(\lambda)$}  &       {$I(\lambda)$}  &       {$\delta I$}     \\
\hline
$E(B-V)$          &             &               &       0.03    &               &               &       0.00    &               &               &       0.13    &               &               &       0.06    &               \\
$\rm H\gamma$   &       4343    &               &               &               &       54.0    &       54.0    &       2.3     &       37.9    &       40.6    &       1.3     &       41.9    &       43.1    &       12.2    \\
{[}O\,{\sc iii}{]}      &       4363    &       2.4     &       2.4     &       0.2     &       8.4     &       8.4     &       0.9     &               &               &               &       12.3    &       12.6    &       3.6     \\
He\,{\sc i}     &       4388    &       0.6     &       0.6     &       0.1     &               &               &               &               &               &               &               &               &               \\
He\,{\sc i}     &       4471    &       4.6     &       4.7     &       0.2     &       3.0     &       3.0     &       0.5     &       3.3     &       3.5     &       0.2     &       6.1     &       6.3     &       1.8     \\
He\,{\sc ii}    &       4541    &               &               &               &               &               &               &               &               &               &       1.9     &       1.9     &       0.6     \\
C\,{\sc iv}     &       4650    &       2.0$^a$ &       2.0$^a$ &       0.2$^a$ &               &               &               &               &               &               &       1.3     &       1.3     &       0.4     \\
He\,{\sc ii}    &       4686    &       0.4     &       0.4     &       0.1     &               &               &               &               &               &               &       58.4    &       59.0    &       16.7    \\
{[}Ar\,{\sc iv}{]}/He{\sc i}    &       4712    &       0.7     &       0.8     &       0.1     &       1.6     &       1.6     &       0.4     &       0.8     &       0.8     &       0.2     &       5.3     &       5.4     &       1.6     \\
{[}Ne\,{\sc iv}{]}      &       4726    &               &               &               &               &               &               &               &               &               &       1.5     &       1.5     &       0.5     \\
{[}Ar\,{\sc iv}{]}      &       4741    &               &               &               &               &               &               &               &               &               &       4.9     &       4.9     &       1.4     \\
$\rm H\beta$    &       4861    &       100.0   &       100.0   &       3.6     &       100.0   &       100.0   &       3.7     &       100.0   &       101.1   &       2.9     &       100.0   &       100.0   &       28.3    \\
He\,{\sc i}     &       4922    &       0.8     &       0.8     &       0.1     &       0.6     &       0.6     &       0.3     &       1.0     &       1.0     &       0.1     &       1.4     &       1.4     &       0.4     \\
{[}O\,{\sc iii}{]}      &       4931    &               &               &               &               &               &               &               &               &               &               &               &               \\
{[}O\,{\sc iii}{]}      &       4959    &       101.5   &       101.3   &       3.6     &       197.2   &       197.2   &       7.1     &       57.7    &       57.7    &       1.7     &       89.8    &       89.3    &       25.3    \\
{[}O\,{\sc iii}{]}      &       5007    &       304.5   &       303.4   &       10.8    &       591.5   &       591.5   &       21.1    &       171.7   &       170.7   &       4.9     &       267.4   &       265.3   &       75.1    \\
{[}N\,{\sc i}{]}        &       5197    &               &               &               &               &               &               &               &               &               &       2.2     &       2.1     &       0.7     \\
He\,{\sc ii}    &       5412    &               &               &               &               &               &               &               &               &               &       3.9     &       3.8     &       1.1     \\
{[}Cl\,{\sc iii}{]}     &       5537    &               &               &               &               &               &               &       0.2     &       0.1     &       0.1     &               &               &               \\
C\,{\sc iii}    &       5696    &       0.3$^a$ &       0.3$^a$ &       0.1$^a$ &               &               &               &               &               &               &               &               &               \\
{[}N\,{\sc ii}{]}       &       5755    &       1.1     &       1.1     &       0.1     &               &               &               &       0.8     &       0.7     &       0.1     &       8.5     &       8.1     &       2.4     \\
He\,{\sc i}     &       5876    &       14.1    &       13.8    &       0.5     &       14.6    &       14.6    &       0.6     &       12.7    &       11.7    &       0.4     &               &               &               \\
{[}O\,{\sc i}{]}        &       6302    &       1.8     &       1.8     &       0.1     &               &               &               &       1.6     &       1.4     &       0.1     &       3.1     &       2.9     &       0.9     \\
{[}S\,{\sc iii}{]}      &       6312    &       0.7     &       0.7     &       0.1     &               &               &               &       0.5     &       0.5     &       0.1     &       2.9     &       2.7     &       0.8     \\
{[}O\,{\sc i}{]}        &       6347    &       0.1     &       0.1     &       0.1     &               &               &               &               &               &               &               &               &               \\
{[}O\,{\sc i}{]}        &       6365    &       0.6     &       0.6     &       0.1     &               &               &               &       0.6     &       0.6     &       0.1     &       0.9     &       0.9     &       0.3     \\
{[}N\,{\sc ii}{]}       &       6548    &       8.4     &       8.2     &       0.3     &               &               &               &       14.3    &       12.5    &       0.4     &       62.1    &       58.0    &       16.4    \\
$\rm H\alpha$   &       6563    &       295.5   &       286.0   &       10.2    &       270.4   &       270.4   &       9.6     &       327.2   &       286.0   &       8.1     &       306.5   &       286.0   &       80.9    \\
{[}N\,{\sc ii}{]}       &       6584    &       22.9    &       22.1    &       0.8     &       1.4     &       1.4     &       0.2     &       41.3    &       36.0    &       1.1     &       174.2   &       162.4   &       46.0    \\
He\,{\sc i}     &       6678    &       3.6     &       3.5     &       0.2     &       3.7     &       3.7     &       0.2     &       3.6     &       3.1     &       0.1     &       4.9     &       4.5     &       1.3     \\
{[}S\,{\sc ii}{]}       &       6716    &       0.3     &       0.3     &       0.1     &               &               &               &       1.3     &       1.1     &       0.1     &       3.9     &       3.6     &       1.1     \\
{[}S\,{\sc ii}{]}       &       6730    &       0.6     &       0.6     &       0.1     &               &               &               &       2.5     &       2.1     &       0.1     &       6.8     &       6.3     &       1.8     \\
He\,{\sc i}     &       7002    &       0.2     &       0.2     &       0.1     &               &               &               &               &               &               &               &               &               \\
{[}Ar\,{\sc v}{]}       &       7007    &               &               &               &               &               &               &               &               &               &       2.2     &       2.0     &       0.6     \\
He\,{\sc i}     &       7065    &       10.3    &       9.9     &       0.4     &       3.9     &       3.9     &       0.3     &       8.2     &       7.0     &       0.3     &       12.5    &       11.5    &       3.3     \\
{[}Ar\,{\sc iii}{]}     &       7100    &       0.3     &       0.3     &       0.1     &               &               &               &               &               &               &               &               &               \\
{[}Ar\,{\sc iii}{]}     &       7136    &       5.2     &       5.0     &       0.2     &       1.9     &       1.9     &       0.2     &       3.4     &       2.8     &       0.1     &       9.1     &       8.4     &       2.4     \\
{[}Ar\,{\sc iv}{]}      &       7172    &               &               &               &               &               &               &               &               &               &       1.4     &       1.3     &       0.4     \\
{[}Ar\,{\sc iv}{]}      &       7237    &       1.7     &       1.6     &       0.1     &               &               &               &       0.8     &       0.7     &       0.1     &               &               &               \\
He\,{\sc i}     &       7281    &       1.0     &       1.0     &       0.1     &               &               &               &       1.0     &       0.9     &       0.1     &       1.3     &       1.2     &       0.4     \\
{[}O\,{\sc ii}{]}       &       7320    &       16.4    &       15.7    &       0.6     &               &               &               &       13.2    &       11.1    &       0.4     &       4.2     &       3.8     &       1.1     \\
{[}O\,{\sc ii}{]}       &       7330    &       13.3    &       12.7    &       0.5     &               &               &               &       10.2    &       8.5     &       0.3     &       3.4     &       3.1     &       0.9     \\
\hline
$^a$ stellar line
%\end{tabular}
\end{longtable}

%%%%%%%%%%%%%%%%%%%%%%%%%%%%%%%%%%%%%%%%%%

\begin{table}
\begin{center}
\caption{The modeled and observed line-flux ratios in the five MC PNe showing the evolution of the [O\,{\sc iii}] 5007\,\AA\ line flux.}\label{tab:models}
\begin{tabular}{*{20}{c}}
\hline
\hline
        &               &       \multicolumn{2}{c}{MGPN LMC 39}         &       \multicolumn{2}{c}{SMP LMC 31}         &       \multicolumn{2}{c}{SMP LMC 55}  &       \multicolumn{2}{c}{MGPN SMC 8}  &       \multicolumn{2}{c}{SMP SMC 1}           \\
ID      &       $\lambda$       &       Model   &       Observed        &       Model   &       Observed        &       Model   &       Observed        &       Model   &       Observed        &       Model   &       Observed        \\ \hline
{[}O\,{\sc iii}]        &       4363    &               &               &       0.8     &       1.1     &       1.4     &       1.5     &       1.3     &       2.6     &       2.5     &       2.4     \\
He\,{\sc i}     &       4471    &       2.1     &       3.1     &       1.7     &       3.2     &       2.9     &       3.7     &       2.7     &       4.8     &       3.4     &       4.7     \\
He\,{\sc i}     &       4713    &               &       0.6     &       0.2     &       0.5     &       0.4     &       0.6     &               &               &       0.5     &       0.8     \\
$\rm H\beta$    &       4861    &       100.0   &       100.0   &       100.0   &       100.0   &       100.0   &       100     &       100.0   &       100.0   &       100.0   &       100.0   \\
He\,{\sc i}     &       4922    &               &               &       0.5     &       0.9     &       0.8     &       0.8     &       0.7     &       1.0     &       0.9     &       0.8     \\
{[}O\,{\sc iii}]        &       4959    &       47.9    &       52.5    &       42.0    &       47.1    &       58.7    &       62.9    &       63.5    &       52.5    &       99.2    &       101.5   \\
{[}O\,{\sc iii}]        &       5006    &       143.0   &       160.1   &       125.4   &       138.6   &       175.1   &       187.3   &       189.6   &       162.5   &       296.0   &       303.4   \\
{[}N\,{\sc i}]  &       5197    &               &               &       0.0     &       0.3     &       1.8     &       0.4     &               &               &               &               \\
{[}N\,{\sc ii}] &       5755    &       1.0     &       1.6     &       0.6     &       2.8     &       8.4     &       1.9     &               &               &       1.9     &       1.1     \\
He\,{\sc i}     &       5876    &       6.0     &       8.7     &       4.3     &       10.3    &               &               &               &               &               &               \\
{[}O\,{\sc i}]  &       6300    &       3.8     &       1.6     &       1.6     &       2.2     &       3.1     &       2.2     &       4.2     &       5.2     &       2.9     &       1.8     \\
{[}S\,{\sc iii}]        &       6312    &       2.8     &       1.3     &       0.6     &       0.7     &       0.7     &       0.9     &       0.7     &       0.9     &       0.2     &       0.7     \\
{[}O\,{\sc i}]  &       6363    &       1.2     &       0.4     &       0.5     &       0.8     &       1.0     &       0.7     &       1.3     &       1.7     &       0.9     &       0.6     \\
{[}N\,{\sc ii}] &       6548    &       8.0     &       9.3     &       12.5    &       15.5    &       19.9    &       25.4    &       30.2    &       27.6    &       24.1    &       8.2     \\
$\rm H\alpha$   &       6563    &       277.8   &       286.0   &       277.2   &       286.0   &       277.4   &       286.0   &       278.3   &       286.0   &       277.7   &       286.0   \\
{[}N\,{\sc ii}] &       6584    &       23.7    &       25.3    &       36.8    &       46.6    &       58.8    &       75.9    &       88.9    &       86.0    &       71.0    &       22.1    \\
He\,{\sc i}     &       6678    &       1.6     &       2.3     &       1.3     &       2.8     &       2.2     &       2.9     &       2.0     &       0.7     &       2.6     &       3.5     \\
{[}S\,{\sc ii}] &       6716    &       2.2     &       0.6     &       3.4     &       0.3     &       1.0     &       1.2     &       6.9     &       7.8     &       0.4     &       0.3     \\
{[}S\,{\sc ii}] &       6731    &               &               &       4.8     &       0.7     &       1.8     &       2.4     &       9.4     &       10.5    &       0.7     &       0.6     \\
He\,{\sc i}     &       7065    &               &               &       1.5     &       6.9     &       3.9     &       6.9     &       2.4     &       4.7     &       5.0     &       9.9     \\
{[}Ar\,{\sc iii}]       &       7136    &               &               &       1.9     &       5.2     &       5.1     &       6.3     &       6.2     &       4.6     &       5.9     &       5.0     \\
{[}O\,{\sc ii}] &       7320    &       26.6    &       33.5    &       10.3    &       27.0    &       20.2    &       19.6    &       8.6     &       10.8    &       11.9    &       15.7    \\
{[}O\,{\sc ii}] &       7330    &       14.3    &       24.9    &       8.4     &       21.2    &       10.9    &       14.9    &       7.0     &       8.2     &       6.4     &       12.7    \\
\end{tabular}
\end{center}
\end{table}
\end{landscape}
\end{appendix}

\end{document}